\documentclass[twocolumn]{aastex62}
\usepackage{amsmath}

\def\beq{\begin{eqnarray}}
\def\eeq{\end{eqnarray}}

\submitjournal{ApJ}
\shorttitle{Black hole hyperaccretion model for $\rm ^{56}Ni$ bumps}
\shortauthors{Song \& Liu}

\begin{document}

\title{Black hole hyperaccretion inflow-outflow model. II. Long-duration gamma-ray bursts and supernova $\rm ^{56}Ni$ bumps}

\correspondingauthor{Tong Liu}
\email{tongliu@xmu.edu.cn}

\author{Cui-Ying Song}
\affiliation{Department of Astronomy, Xiamen University, Xiamen, Fujian 361005, China}

\author{Tong Liu}
\affiliation{Department of Astronomy, Xiamen University, Xiamen, Fujian 361005, China}

\begin{abstract}
Long-duration gamma-ray bursts (LGRBs) associated with supernovae (SNe) are possibly born out of the death of a massive star. After the star collapses, a stellar-mass black hole (BH) is formed, surrounded by a hyperaccretion disk with outflows. Blandford-Znajek jets can be launched and then break out from the envelope to power LGRBs. The jet luminosity depends on the net inflow accretion rate at the inner radius of the disk. Furthermore, $\rm ^{56}Ni$ synthesis should occur in the strong outflows from the accretion disk. The decay of $\rm ^{56}Ni$ is considered to be the possible origin of SN bumps in the subsequent optical afterglows of LGRBs. If $\rm ^{56}Ni$ originates entirely from the outflows, there is competition between the luminosities of LGRBs and those of the corresponding $\rm ^{56}Ni$ bumps because of the material distribution between the disk inflows and outflows. In this paper, we investigated these two luminosities based on 15 cases of LGRB-SN in the framework of the BH hyperaccretion inflow-outflow model. Then, one can constrain the characteristics of the progenitor stars of these LGRBs. The results indicate that these LGRBs may originate from the low-metallicity ($Z\lesssim 10^{-2}Z_{\odot}$, where $Z$ and $Z_{\odot}$ are the metallicities of the stars and the Sun, respectively) stars or some massive solar-metallicity stars. For ultra-LGRBs (ULGRBs), such as GRB 111209A, most of the massive low-metallicity stars with $Z \lesssim 10^{-2}Z_{\odot}$ could be progenitors only if very strong outflows are launched from the disks. When the contributions of nucleosynthesis in the disk outflows are considered, there is no shortage of $\rm ^{56}Ni$ mass for luminous SNe associated with ULGRBs.
\end{abstract}

\keywords{accretion, accretion disks - black hole physics - gamma-ray burst: general - nuclear reactions, nucleosynthesis, abundances - supernovae: general}

\section{Introduction}\label{sec:intro}

As the most luminous sources in the universe, gamma-ray bursts (GRBs) can typically release $\sim10^{53}-10^{54}~\rm erg$ of isotropic energy within seconds. Based on the GRB duration, $T_{90}$ \citep{Kouveliotou1993}, they are classified as short-duration GRBs ($T_{90}< 2~\rm s$, SGRBs) and long-duration GRBs ($T_{90}> 2 ~\rm s$, LGRBs). It was proposed that the two types correspond to physically distinct progenitors \citep[e.g.,][]{Eichler1989,Paczynski1991,Narayan1992,Woosley1993}.

In recent decades, multi-wavelength observations of advanced space-based and ground instruments have increased our understanding of the progenitors and central engines of GRBs. These observations have indicated that some LGRBs are associated with broad-line type Ib/c SNe \citep[e.g.,][]{Woosley2006,Cano2016,Guessoum2017}. The connection between GRB 980425 and SN 1998bw have provided the first clue regarding association of LGRBs with SNe \citep[]{Galama1998}. The SN had a very large kinetic energy of $\sim2-5\times10^{52} ~\rm erg$ and occurred nearly simultaneously with the GRB. However, the gamma-ray luminosity of GRB 980425 ($L_{\rm \gamma,\rm iso}\sim5\times10^{46}~\rm erg~s^{-1}$) was more than three orders of magnitude fainter than that of typical LGRBs \citep[e.g.,][]{Frail2001,Bloom2003}, which is not sufficient evidence of a physical connection. A compelling spectroscopic association between high-luminosity GRB 030329 ($L_{\rm \gamma,\rm iso}\sim8\times10^{50}~\rm erg~s^{-1}$) and SN 2003dh have provided conclusive evidence \citep[e.g.,][]{Hjorth2003,Matheson2003,Stanek2003}. More LGRB-SN cases have been discovered since the launch of the \emph{Swift} satellite. Other significant evidence \citep[e.g.,][]{Zhang2007}, such as host galaxies, suggesting that LGRBs possibly originate from collapsars \citep[e.g.,][]{Woosley1993,MacFadyen1999}. Furthermore, some GRBs with extremely long durations $\sim 10^{4} \rm s$, known as ultra-LGRB \citep[ULGRBs, e.g.,][]{Levan2014}, have been observed. Some ULGRBs associated with SNe were also discovered. Thus, they might be produced by the collapsars \citep[e.g.,][]{Liu2018a,Liu2018b}.

A stellar-mass black hole (BH) surrounded by an accretion disk \citep[e.g.,][]{Woosley1993,MacFadyen1999,Popham1999} or a rapidly spinning (period $\sim 1~\rm ms$), highly magnetized (surface magnetic field $\rm \sim 10^{15} ~G$) neutron star \citep[magnetar, see e.g.,][]{Usov1992,Wheeler2000} might be produced in the center of the GRB progenitors. The spin-down of magnetars can power GRBs, as has been widely studied in recent years \citep[e.g.,][]{Bucciantini2008,Bucciantini2009,Metzger2011,Metzger2018,Cano2016,Yu2017}.

In the BH scenario, the accretion disk is in a hyperaccretion phase because of its high rate \citep[$\gtrsim 10^{-8}~M_{\odot}~\rm s^{-1}$, see e.g.,][]{Liu2018a,Liu2018b}. Energy is released by the neutrino radiation process or the Blandford-Znajek \citep[hereafter BZ,][]{Blandford1977} mechanism by extracting the gravitational or rotational energy of the central BH, respectively. Generally, the BZ mechanism is more effective than the neutrino annihilation process to power GRBs \citep[e.g.,][]{Liu2015}. If the neutrino radiation is dominated by cooling, the disk is referred to as a neutrino-dominated accretion flow \citep[NDAF, e.g.,][]{Popham1999,Liu2017,Liu2018a}. Once the massive outflows escape from the disk, the net inflow accretion rate in the inner region may be too low ($\lesssim 10^{-3}~M_{\odot}~\rm s^{-1}$) to effectively produce neutrinos; then, the BZ mechanism becomes dominant in the rotating BH scenario \citep[e.g.,][]{Liu2018a,Liu2018b}. Of course, the BZ mechanism can dominantly power GRBs for the high-accretion-rate cases. Bipolar relativistic jets are launched through the BZ mechanism or the neutrino annihilation mechanism \citep[e.g.,][]{Popham1999,Di2002,Liu2007,Liu2014,Liu2017}. The jets travel through the interior of the progenitor, interacting with the in-falling materials. If the jet break out, an energetic GRB is generated. For the BH hyperaccretion inflow-outflow model in the core-collapsar scenario, the BZ jets are naturally sufficient to power LGRBs \citep[e.g.,][]{Liu2018b}.

The light curves of SNe are mainly driven by the decay of radioactive $\rm ^{56}Ni$, and its daughter $\rm ^{56}Co$ to $\rm ^{56}Fe$. The half-lives of these decays are 6.077 days and 77.236 days, respectively \citep[e.g.,][]{Arnett1982,Woosley1986}. In these radioactive decay processes, gamma-ray photons are emitted, which then thermalize in the SN ejecta. The optically thick ejecta are heated and then radiate energy from the decays in the optical and near-infrared bands. Thus, the $\rm ^{56}Ni$ mass is closely linked with the luminosity of SNe. Statistical analysis of the bolometric properties of SNe shows that the average $^{56}\rm Ni$ mass is $0.4\pm0.2~M_{\odot}$ in an explosion \citep[e.g.,][]{Cano2017}. Additionally, during the explosive burning of the collapsar, $\rm ^{56}Ni$ synthesis also occurs in the hyperaccretion disk \citep[e.g.,][]{Chakrabarti1987,Surman2008,Liu2013} or the winds/outflows from the accretion disk \citep[e.g.,][]{Pruet2004,Kohri2005,Surman2005,Surman2006,Surman2011,Hu2008,Liu2013,Hu2015,Wu2016}. This disk is widely considered the main factory of $\rm ^{56}Ni$ in these studies. Moreover, \cite{Suwa2015} investigated the amount of $\rm ^{56}Ni$ produced by a rapidly spinning magnetar. They found that the $\rm ^{56}Ni$ mass depends on the strength of the initial angular velocity and the dipole magnetic field.

This paper is the second work in a series on the BH hyperaccretion inflow-outflow model. In paper I \citep{Liu2018b}, we studied the masses and metallicities of the progenitor stars of LGRBs and ULGRBs in the collapsar scenario combined with GRB observations. The results show that LGRBs lasting from several seconds to tens of seconds in the rest frame can be produced only by some zero-metallicity stars or solar-metallicity ($Z\sim 1~Z_{\odot}$, where $Z$ and $Z_{\odot}$ are the metallicities of progenitor stars and the Sun), massive ($M\geq34~M_{\odot}$, where $M$ and $M_{\odot}$ are the masses of progenitor stars and the Sun) stars. ULGRBs, such as GRB 111209A, may originate from a fraction of the low-metallicity ($Z\leq10^{-2}~Z_{\odot}$) stars or Population III stars. The fraction of LGRBs lasting less than tens of seconds in the rest frame is obviously larger than the fraction of the progenitor stars of interest. This finding compels us to believe that the authentic activity timescale of the central engine should be longer than the timescale of prompt emission.

The disk inflows and outflows compete for the materials in the envelope. The luminosities of LGRBs and those of their $\rm ^{56}Ni$ bumps are determined by the inflows and outflows, respectively. Thus the competition in terms of masses and energies between LGRBs and SN bumps can constrain the natures of the progenitor stars. In this paper, we further studied the characteristics of the progenitors of LGRBs and ULGRBs by using the BH hyperaccretion inflow-outflow model and the observational data of LGRBs-SNe. Our model is described in Section 2. The results are shown in Section 3. We summarize the conclusions in Section 4.

\section{Model}

\subsection{Jet luminosity}

After a massive star collapses, a stellar-mass rotating BH surrounded by a hyperaccreting disk might form. As shown in paper I \citep{Liu2018b}, the outflows in this system play important roles \citep[e.g.,][]{Yuan2012,Yuan2014,Sadowski2015}. The accretion rate at the outer boundary of the disk is determined by the mass supply from a certain progenitor star. Here, we define the dimensionless factor $f$, the fraction of the outflow mass rate to the mass supply rate from the envelope $\dot{M}_{\rm pro}$, to parameterize the effect of the outflows. The net accretion rate at the inner radius of the disk $\dot{M}_{\rm inflow}$ can be expressed by
\beq
\dot{M}_{\rm inflow}=\dot{M}_{\rm pro}(1-f).
\eeq
The mass supply rate can be determined by the density $\rho$ profile and the mass coordinate $M_{r}$ in the pre-SN model \citep[e.g.,][]{Suwa2011,Woosley2012},
\beq
\dot{M}_{\rm pro}=\frac{2M_{r}}{t_{\rm ff}(r)}\frac{\rho}{\bar{\rho}-\rho},
\eeq
where $\bar{\rho}=3M_{r}/(4 \pi r^{3})$ is the mean density of the progenitor star and $t_{\rm ff}=\sqrt{3\pi/32G\bar{\rho}}$ is the free fall timescale.

Assuming that the jets are powered by the BZ mechanism, the BZ jet power can be estimated as
\citep[e.g.,][]{Lee2000a,Lee2000b,McKinney2005,Barkov2008,Barkov2010,Komissarov2009,Luo2013,Lei2013,Lei2017,Liu2015}
\beq
L_{\rm BZ}=1.7\times10^{50}a_{*}^{2}m_{\rm BH}^{2}B_{\rm BH,15}^{2}F(a_*){\rm~erg~s^{-1}},
\eeq
where $m_{\rm BH}=M_{\rm BH}/M_\odot$ is the dimensionless mass of the BH; $B_{\rm BH,15}=B_{\rm BH}/10^{15} {\rm G}$ is the dimensionless magnetic field strength near the horizon, normalized to $10^{15} {\rm G}$; and $a_*$ is the dimensionless spin parameter of the BH. Here, $F(a_*)=[(1+q^{2})/q^{2}][(q+1/q)\arctan(q)-1]$, where $q=a_{*}/(1+\sqrt{1-a_{*}^{2}})$.

\begin{deluxetable*}{ccccccccc}
\tablenum{1}
\tablecaption{(U)LGRB-SN data}
\tablewidth{0pt}
\tabletypesize{\scriptsize}
\tablehead{
 \colhead{GRB} & \colhead{SN} & \colhead{$z$} & \colhead{$T_{90}$} &  \colhead{$\theta_{\rm j}$} & \colhead{$E_{\rm \gamma,\rm iso}$} & \colhead{$E_{\rm k,\rm iso}$}  &  \colhead{$M_{\rm Ni}$} &  \colhead{Ref.}  \\
\nocolhead{GRB} & \nocolhead{(SN)} & \nocolhead{z} & \colhead{(\rm s)} & \colhead{(\rm rad)}& \colhead{($10^{50}$\rm erg)} &\colhead{($10^{51}$\rm erg)} &\colhead{($M_{\odot}$)} & \nocolhead{Ref} }
\decimalcolnumbers
\startdata
LGRBs   \\
\hline
980425  &	   1998bw  &	 0.01  &	23.3   &	0.192               &	$0.00929\pm0.00035$     &	0.0631     &   $0.42\pm0.02$  &	1, 2 \\
011121  &      2001ke  &	0.362  &	47     &	0.157               &	$780\pm210$             &	27         &   $0.35\pm0.01$  &	3, 4\\
021211  &	   2002lt  &	1.004  &	2.8    &	$0.0244\sim0.0768$  &	$112\pm13$              &	40         &   $0.16\pm0.14$  &	4, 5 \\
030329  &	   2003dh  &	0.17   &	22.3   &	0.089               &	133                     &	63.1       &   $0.54\pm0.13$  &	1, 2 \\
031203  &	   2003lw  &	0.1    &	40     &	0.157               &	$1.67_{-0.10}^{+0.04}$  &	1.38       &   $0.57\pm0.04$  &	1, 2 \\	
050525  &	   2005nc  &	0.606  &	8.84   &	0.0551               &	$250\pm43$              &	282        &   $0.24\pm0.02$  &	4, 6, 7 \\
081007  &      2008hw  &	0.53   &	9.01   &	$>0.349$            &	$15_{-3}^{+4}$          &	1.5        &   $0.39\pm0.08$  &	4, 8 \\
091127  &	   2009nz  &	0.48   &	68.7   &	0.096               &	$430\pm30$              &	229        &   $0.33\pm0.01$  &	1, 2 \\
101219B &	   2010ma  &	0.552  &	51     &	$>0.298$            &	$34\pm2$                &	$64\pm35$  &   $0.43\pm0.03$  &	4, 9 \\
120422A &	   2012bz  &	0.283  &	5.35   &	0.2                 &	0.45                    &	$\sim1.2$  &   $0.57\pm0.07$  &	2, 7, 10 \\
130427A &	   2013cq  &	0.3399 &	162.83 &	$>0.0873$           &	$9600\pm40$             &	131        &   $0.28\pm0.02$  &	2, 11 \\
130702A &	   2013dx  &	0.145  &	59     &	0.086               &	$6.4_{-1.0}^{+1.3}$     &	377        &   $0.37\pm0.01$  &	2, 12 \\
130831A &	   2013fu  &	0.479  &	32.5   &	$\geq0.123$         &	$46\pm2$                &	114        &   $0.30\pm0.07$   &	4, 13 \\
140606B &	   iPTF14bfu  &	0.384  &	23.6   &	0.14                &	$34.7\pm0.2$            &	$0.39\sim31.85$    &	$0.42\pm0.17$  &	2, 14 \\																																																																																																																																																																																												\hline
 ULGRB  \\
\hline	
111209A &	   2011kl  &	0.677 &	$\sim15000$   &	$>0.21$   &	$5700\pm700$  &	960    &	$2.27\pm0.64$  &	15, 16 \\																																																																																																																																																																																																																																																																											
\enddata

\emph{\rm References}: \\
(1) \citet{Nemmen2012}; (2) \citet{Toy2016}; (3) \citet{Greiner2003}; (4) \citet{Cano2016}; (5) \citet{Holland2004}; (6) \citet{Zhang2007}; (7) \citet{Ryan2015}; (8) \citet{Jin2013}; (9) \citet{Larsson2015}; (10) \citet{Zhang2012}; (11) \citet{Perley2014}; (12) \citet{Singer2013}; (13) \citet{De2016}; (14) \citet{Cano2015}; (15) \citet{Nakauchi2013}; (16) \citet{Kann2016}.

\end{deluxetable*}

By assuming that the magnetic pressure on the BH horizon is equipartitioned with the ram pressure of the innermost part of the accretion disk, one can obtain the BZ jet power as a function of the dimensionless net mass accretion rate at the inner radius of the disk $\dot{m}_{\rm inflow}=\dot{M}_{\rm inflow}/M_{\odot}$ and the BH spin parameter $a_{*}$ \citep{Liu2018b}, i.e.,
\beq
L_{\rm BZ}=9.3\times10^{53}a_*^{2}\dot{m}_{\rm inflow}X(a_*){\rm~erg~s^{-1}},
\eeq
and
\beq
X(a_*)=F(a_*)/(1+\sqrt{1-a_*^{2}})^{2}.
\eeq
Here, $a_{*}$ is set to 0.86 in our calculations because this value of $a_{*}$  is an equilibrium value of the spin evolution of a BH considering both the
accretion and BZ processes \citep[]{Song2015,Lei2017}.

When the jet moves within the stellar envelope, the balance of pressure is established between the jet head and the stellar envelope. The velocity of the jet head in units of the speed of light can be obtained from \citep[e.g.,][]{Matzner2003,Nakauchi2013}
\beq
\beta_{\rm h}(t)=\frac{1}{1+\tilde{L}(t)^{-1/2}},
\eeq
and
\beq
\tilde{L}(t)\equiv\frac{L_{\rm j}(t-r_{\rm h}/c)}{\pi \theta_{\rm j}^{2}r_{\rm h}^{2}\rho(r_{\rm h}) c^3}.
\eeq
The position of the jet head can be calculated by $r_{h}=\int_{0}^{t}c\beta_{h}dt'$, and $\theta_{\rm j}$ is the half-opening angle of the jet. GRBs will be produced after the jets break out from the progenitor, so we define the break-out time $t_{\rm bo}$ as the moment when the jet head reaches the boundary of the progenitor star.

According to the above equations, we can calculate the theoretical jet luminosity $L_{\rm j,t}$ ($\simeq L_{BZ}$) for the progenitor stars with different masses and metallicities. Notably, we set the theoretical values of $\theta_{\rm j}$ as 0.1 and 0.21 for LGRBs and ULGRBs, respectively, in the cases of all progenitor stars.

On the other hand, the GRB jet power can be estimated by the observational GRBs data \citep[e.g.,][]{Fan2011,Liu2015}, i.e.,
\beq
L_{\rm j}\simeq\frac{(E_{\rm \gamma,\rm iso} + E_{\rm k,\rm iso})(1+z)\theta_{\rm j}^{2}}{2T_{90}},
\eeq
where the isotropic radiated energy and the isotropic kinetic energy of afterglows are denoted by $E_{\rm \gamma,\rm iso}$ and $E_{\rm k,\rm iso}$, respectively, and $z$ and $T_{\rm 90}$ are the redshift and prompt emission duration of GRBs. It is worth noting that the activity timescale of the central engine might be much longer than $T_{\rm 90}$ \footnote{\citet{Lv2014} proposed that a real GRB may be observed as a `short' one if the majority of the emission episode is too faint to be detected above the background. This phenomenon is called the `tip-of-iceberg' effect. In other words, the activity timescale of the GRB central engine may be longer than the observed prompt emission time on account of this effect. To determine the progenitor stars of LGRBs, \citet{Liu2018b} reported that the true duration of the burst is actually longer than $T_{90}$, which is consistent with the X-ray afterglow observations and the related statistical analysis \citep[e.g.,][]{Zhang2014}.}, so the values calculated by Equation (8) should be the upper limit of the LGRB luminosity, as shown in Table 1.

\subsection{$^{56}\rm Ni$ mass}

In light of the photometric and spectroscopic properties, the basic explosion parameters of SNe can be derived using simple analytic models. The Arnett-Valenti relation describes the light curve of Type I SNe \citep[e.g .,][]{Arnett1982,Valenti2008}, i.e.,
\beq
L_{\rm SN}(t)=& M_{\rm Ni}&e^{-x^2}[(\epsilon_{\rm Ni}-\epsilon_{\rm Co})\times \int_{0}^{x}A(k)dk \nonumber
\\ &+&\epsilon_{\rm Co}\int_{0}^{x}B(k)dk],
\eeq
where
\beq
A(z)=2k \exp(-2ky+k^2),
\eeq
\beq
B(z)=2k \exp(-2ky+2ks+k^2),
\eeq
and $x\equiv t/\tau_{\rm m}$, $y\equiv \tau_{\rm m}/(2\tau_{\rm Ni})$, and $s\equiv \tau_{\rm m}(\tau_{\rm Co}-\tau_{\rm Ni})/(2\tau_{\rm Co}\tau_{\rm Ni})$. The decay times of $^{56}\rm Ni$ and $^{56}\rm Co$ are $\tau_{\rm Ni}=8.77 \rm~d$ and $\tau_{\rm Co}=111.3 \rm ~d$, respectively \citep[e.g.,][]{Woosley1986}. The energy produced by one gram of $^{56}\rm Ni$ and $^{56}\rm Co$ in one second was taken as $\epsilon_{\rm Ni}=3.90\times10^{10}~\rm erg~s^{-1} ~g^{-1}$ and $\epsilon_{\rm Co}=6.78\times10^{9}~\rm erg~s^{-1} ~g^{-1}$ \citep[e.g.,][]{Sutherland1984,Cappellaro1997}. From the above equations, one can estimate the $^{56}\rm Ni$ mass $M_{\rm Ni}$ based on the observational SN data, as shown in Table 1.

\begin{figure*}
\centering
\includegraphics[width=7cm,height=7cm]{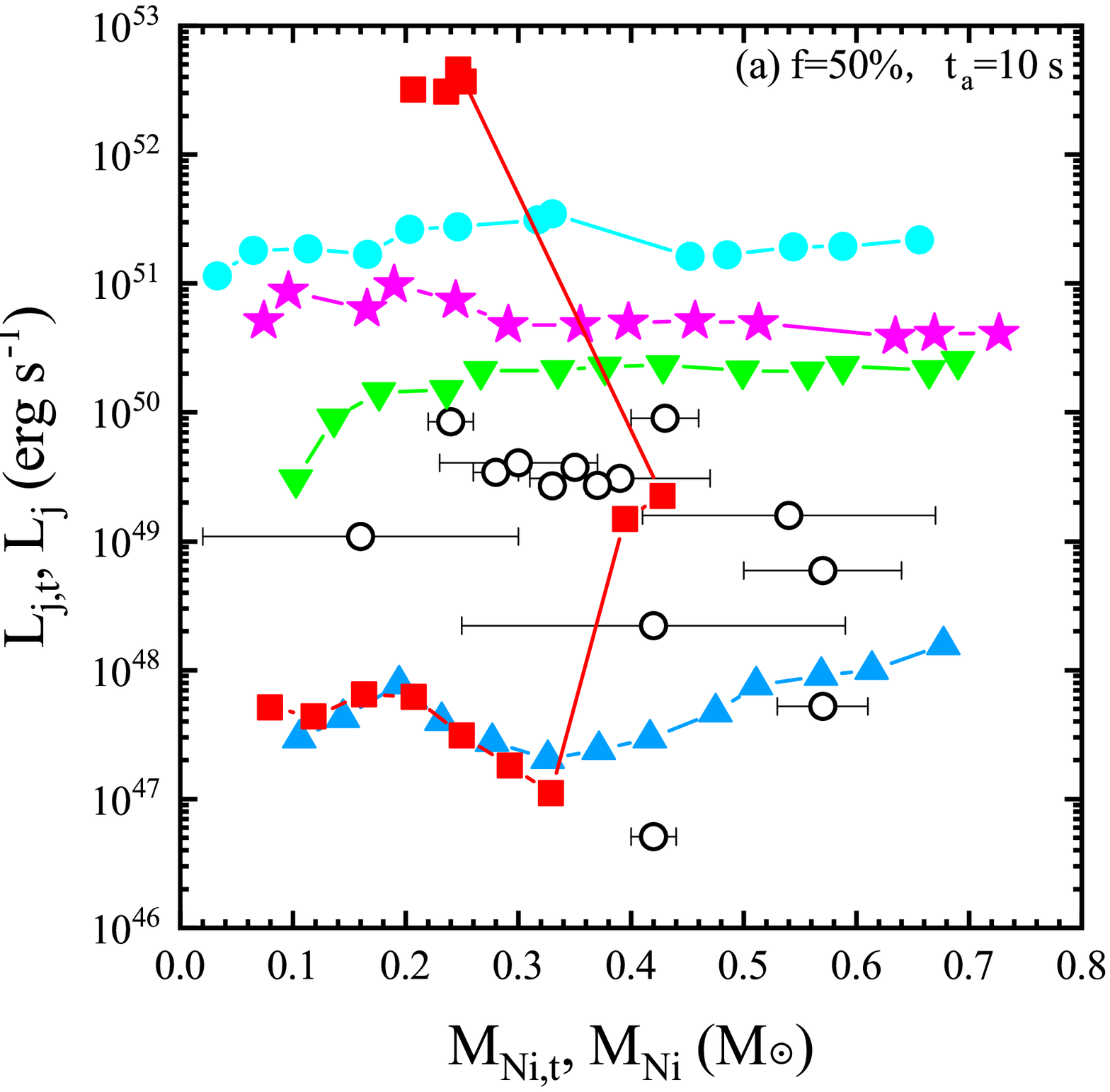}
\includegraphics[width=7cm,height=7cm]{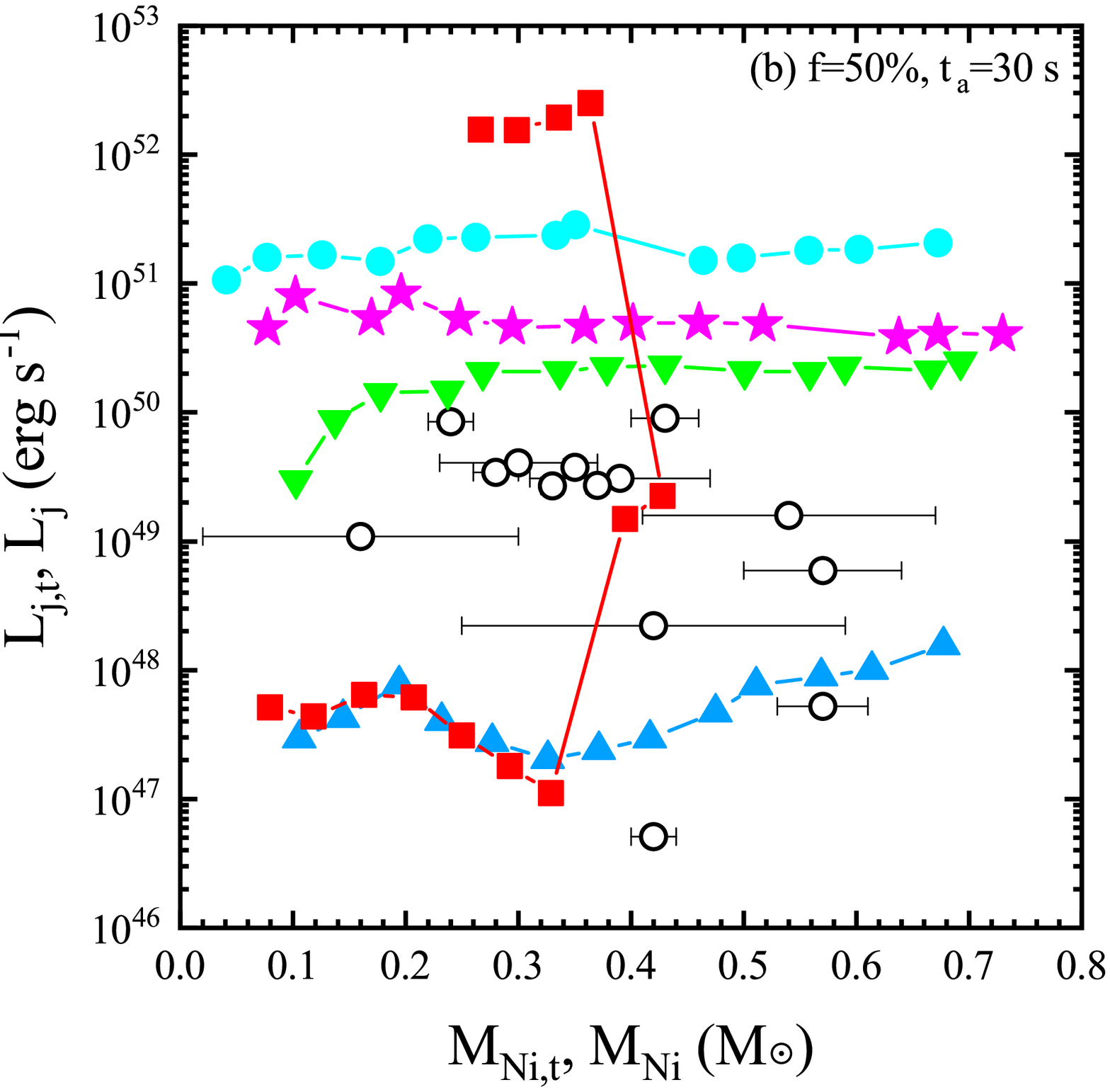}
\includegraphics[width=7cm,height=7cm]{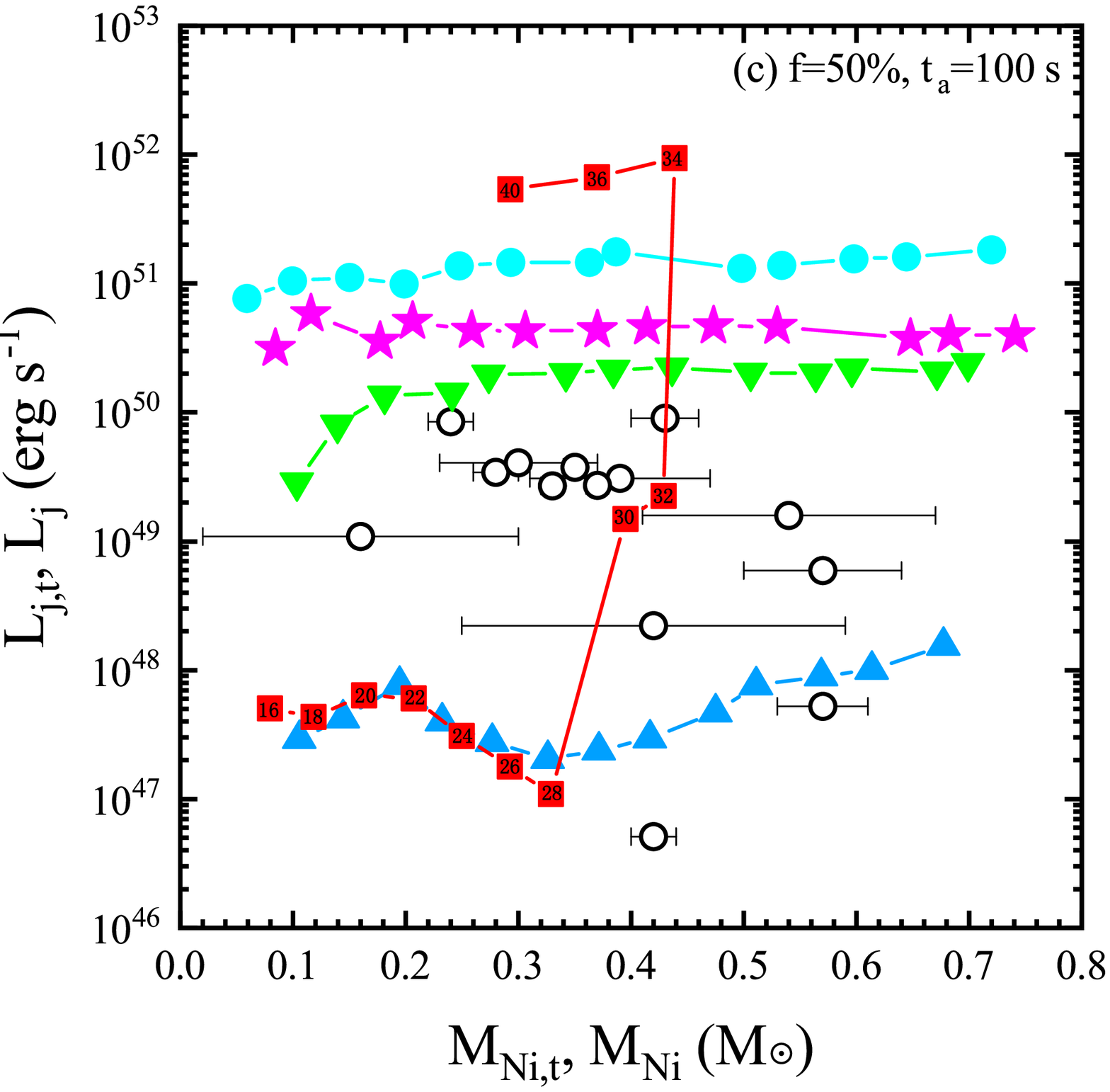}
\includegraphics[width=7cm,height=7cm]{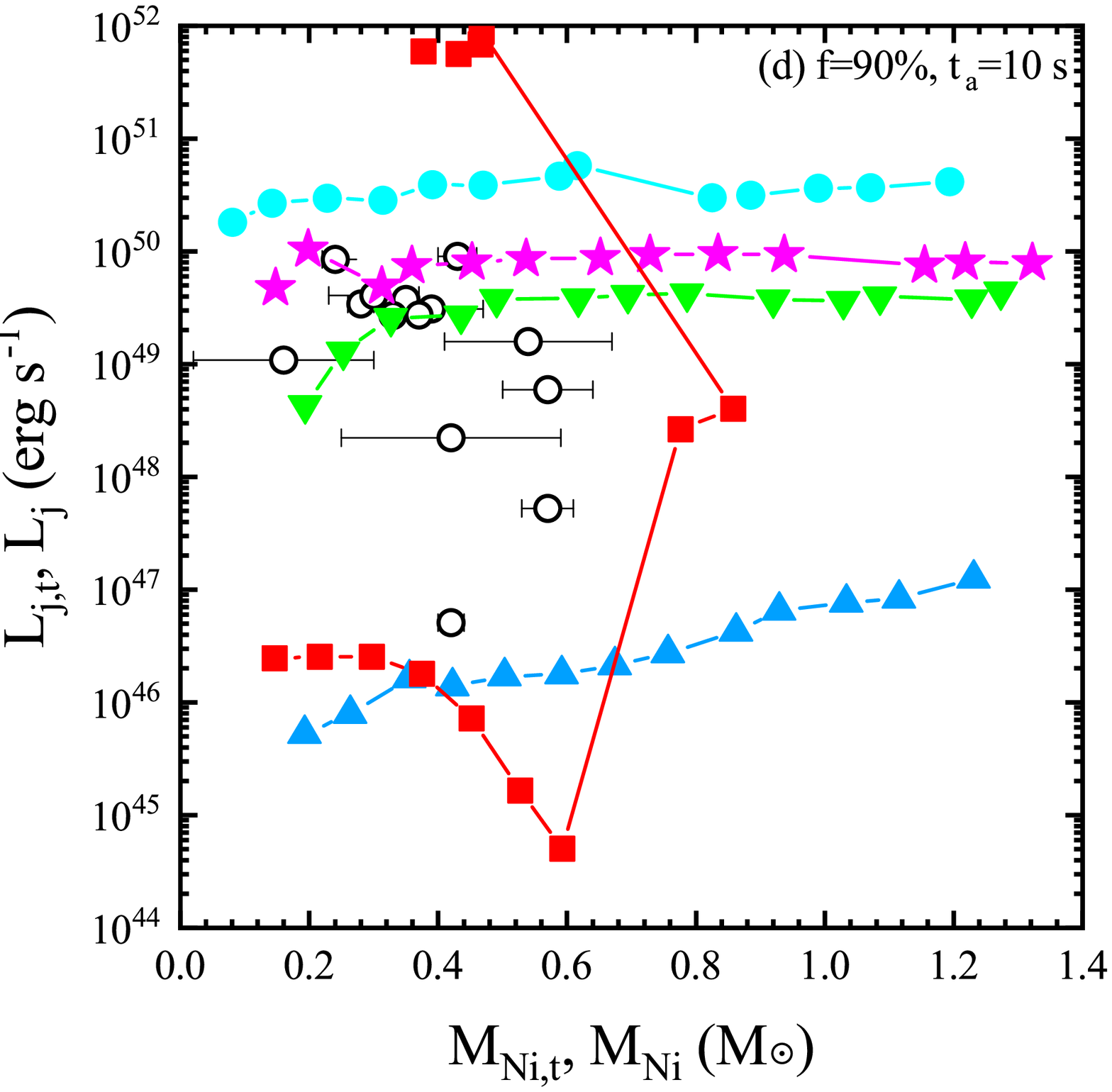}
\includegraphics[width=7cm,height=7cm]{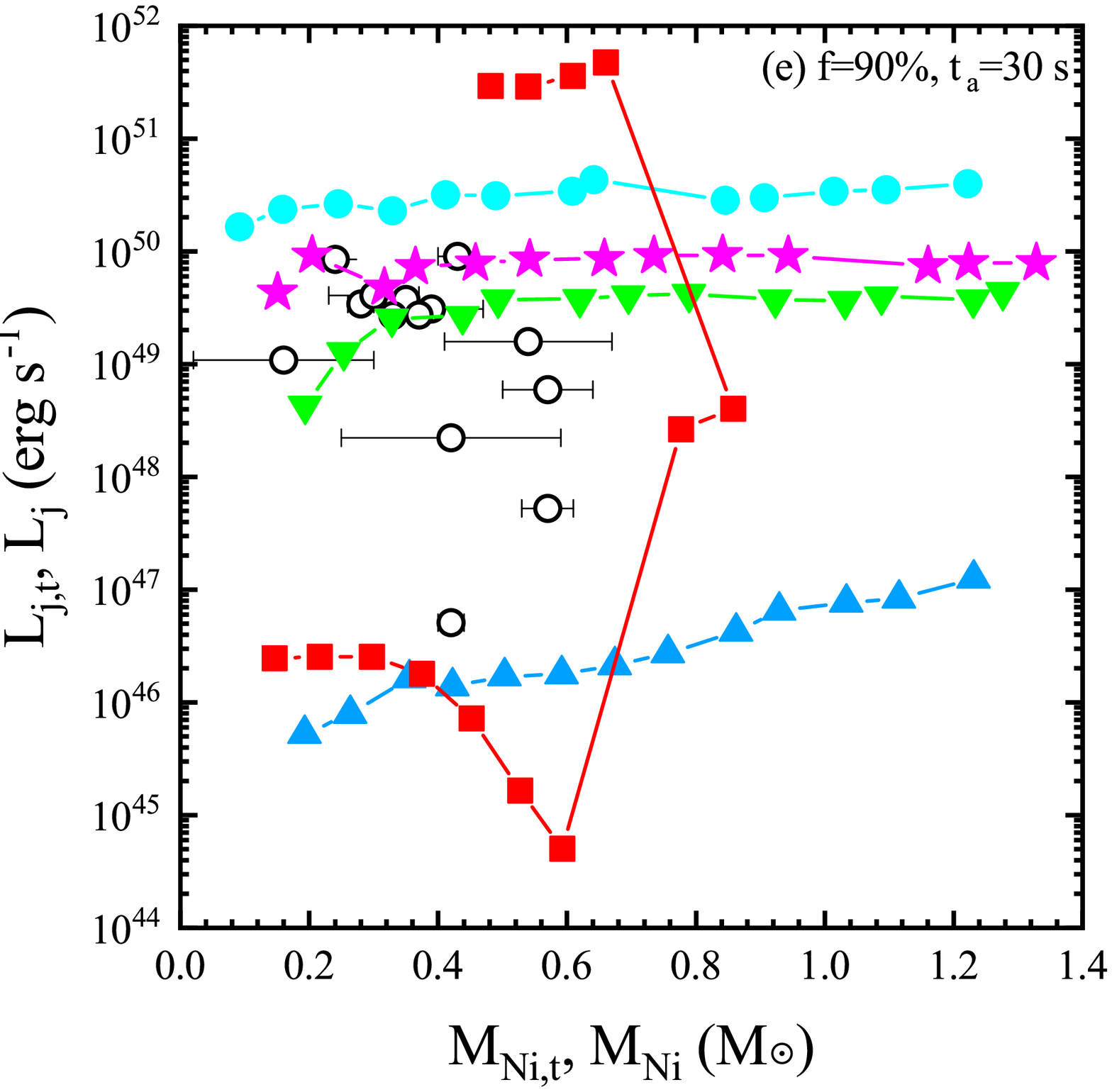}
\includegraphics[width=7cm,height=7cm]{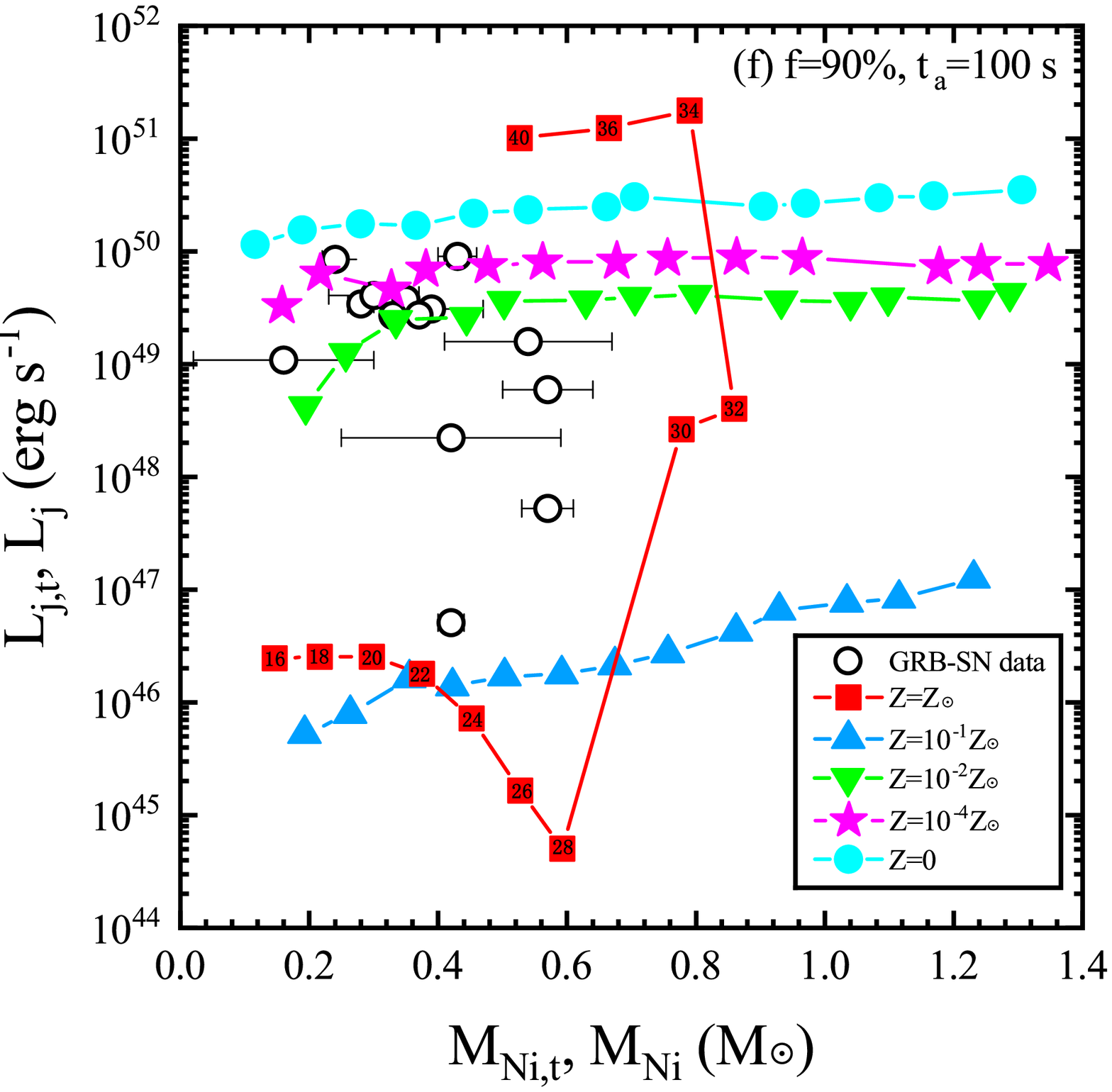}
\caption{Constraints on the masses and metallicities of the progenitor stars of LGRB-SN cases for different outflow rates ($f=50\%$ and $90\%$) and observable activity durations of the LGRB central engine ($t_{\rm a}=10\rm~s$, $30\rm~s$, and $100 \rm~s$). The observational data are denoted by empty black circles. The colored filled symbols indicate the metallicity values $Z=Z_{\odot}, 10^{-1}Z_{\odot}, 10^{-2}Z_{\odot}, 10^{-4}Z_{\odot}$, and 0. Symbols with the same metallicity but different masses are connected by lines. The progenitor mass is in the range of $16-40~M_{\odot}$ with an interval of $2~M_{\odot}$. In Figures (c) and (f), the progenitor stars with 38 $M_\odot$ and $Z_{\odot}$ cannot supply the accretion processes lasting approximately 100 s.}
\end{figure*}

In this paper, we assume that the SN bump is powered purely by $^{56}\rm Ni$ synthesized in the outflows from the disk. Thus, the timescale of the SN light curve $\tau_{\rm m}$ is defined as \citep{Cano2013}
\beq
\tau_{\rm m}\approx(\frac{\kappa}{\beta c})^{1/2}(\frac{M_{\rm outflow}}{v_{\rm ph}})^{1/2},
\eeq
where the integration constant is set as $\beta \approx 13.8$ \citep{Arnett1982} and $v_{\rm ph}$ denotes the peak photospheric velocity, with a typical value of approximately $20,000 \rm~km/s$ \citep{Cano2016}. In addition, we assume a constant opacity $\kappa=0.07~\rm cm^{2} ~g^{-1}$ \citep{Chugai2000}.

It is reasonable that approximately ten percent of the outflow materials are converted  into $^{56}\rm Ni$ through nucleosynthesis \citep[see e.g.,][]{Surman2011}. Then, the theoretical $^{56}\rm Ni$ mass can be estimated by
\beq
M_{\rm Ni,t}\simeq 0.1\int_{0}^{t_{\rm a}+t_{\rm bo}} \dot{M}_{\rm outflow}dt,
\eeq
where $t_{\rm a}$ represent the activity timescale of the central engine after the jets break out from the envelope, as well as the observable activity timescale of the central engine. In other words, $t_{\rm a}+t_{\rm bo}$ is the activity timescale of the GRB central engine. It should also be noted that the outflows are launched after the BH accretion forms, so nucleosynthesis starts at this moment. However, the observable GRBs start when the jet breaks out from the envelope. Since $t_{\rm a}$ should be much longer than $T_{90}$, we take $t_{\rm a}=10\rm~s$, $30\rm~s$, and $100 \rm~s$ in our calculations to demonstrate the rationality of the model.

\section{Results}

\begin{figure*}
\centering
\includegraphics[width=8cm,height=8cm]{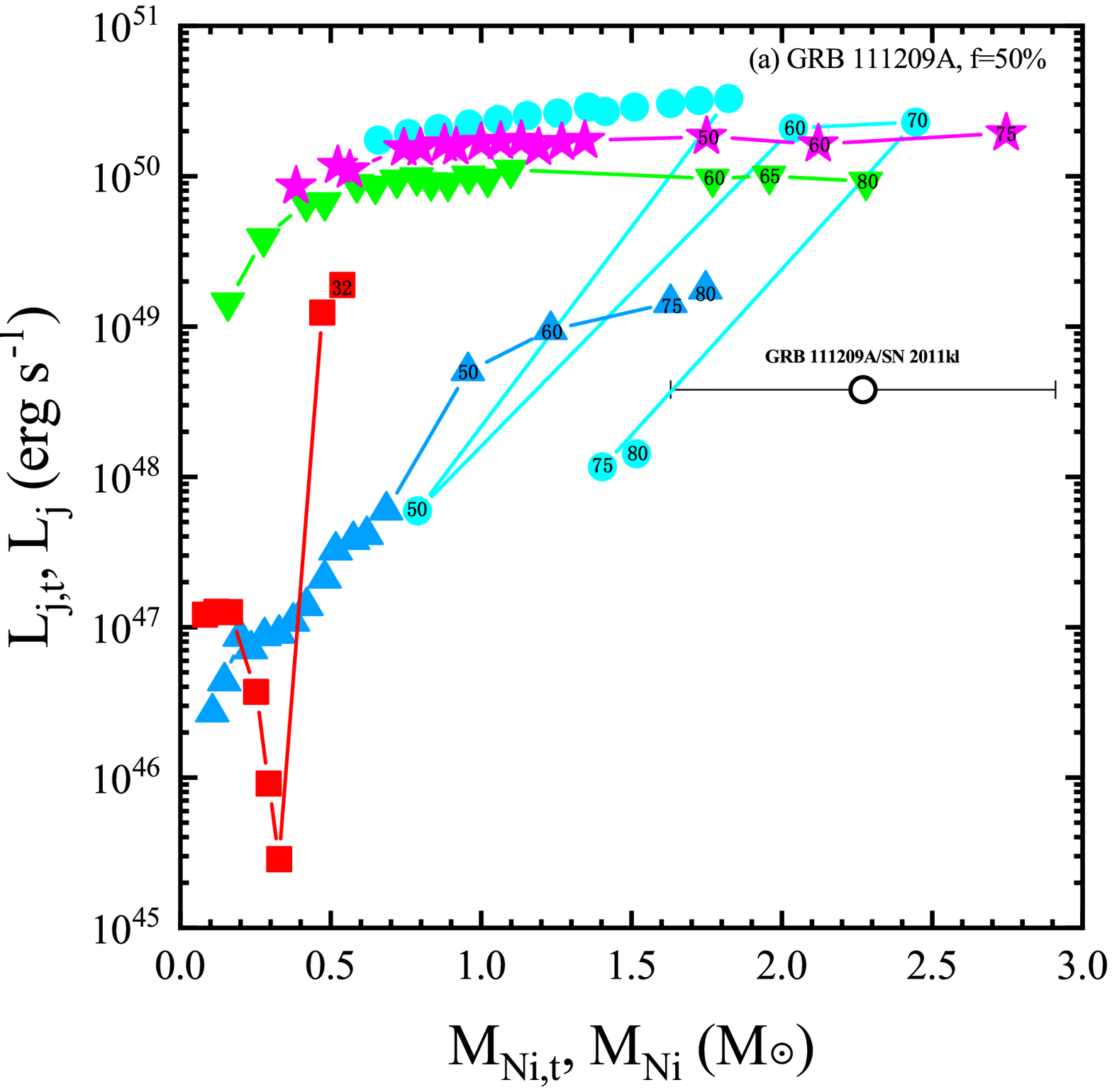}
\includegraphics[width=8cm,height=8cm]{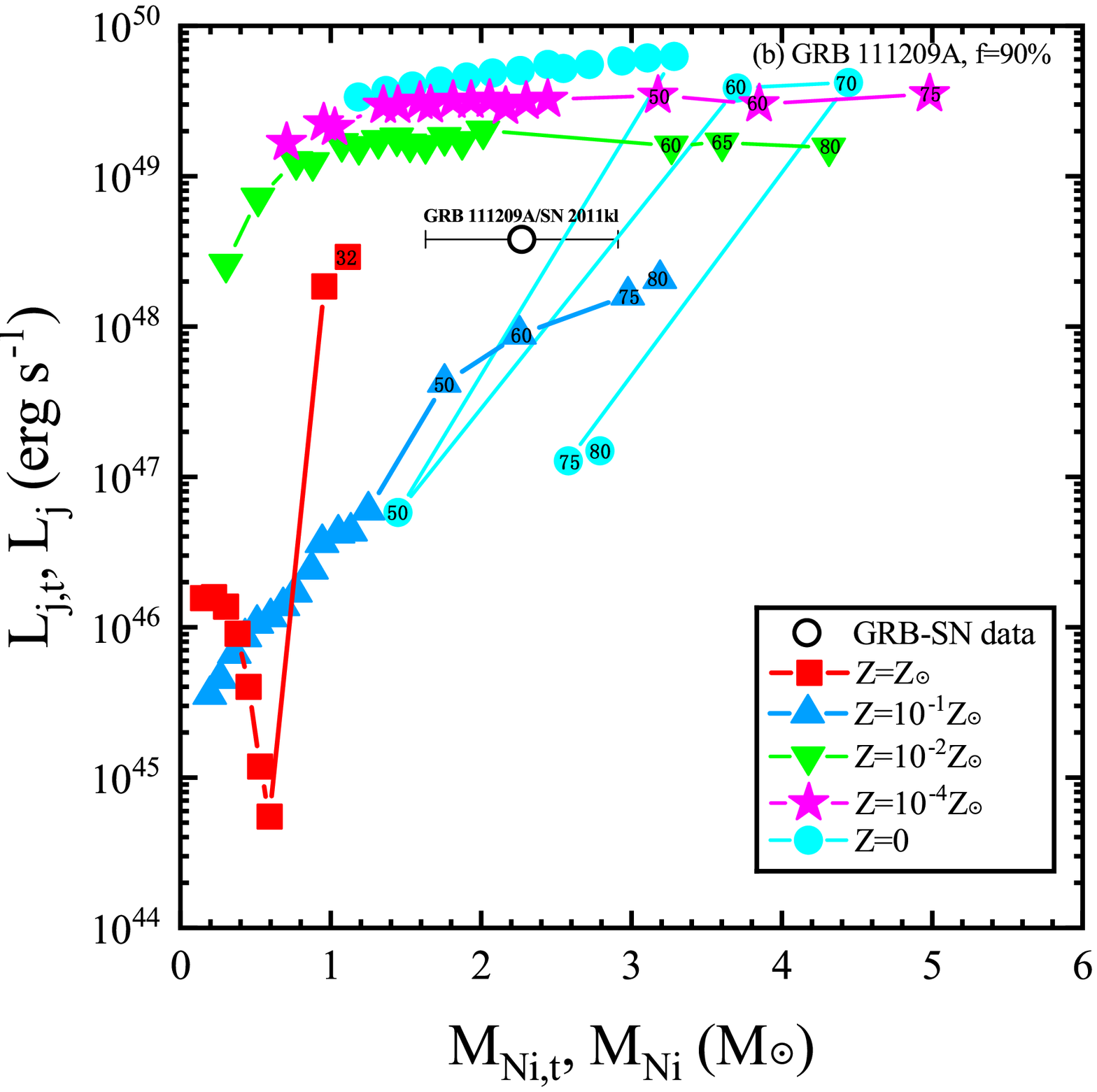}
\caption{Constraints on the masses and metallicities of the progenitor stars of ULGRB GRB 111209A for different outflow rates of $f=50\%$ and $90\%$. The symbols are the same as in Figure 1. We set the progenitor star masses in the range of $16-40~M_{\odot}$ with an interval of $2~M_{\odot}$ and mark the numbers in units of $M_{\odot}$ for mass values larger than $40~M_{\odot}$. In these two figures, the progenitor stars with $> 32~M_\odot$ and $Z_{\odot}$ cannot supply the accretion processes with durations as long as that of GRB 111209A.}
\end{figure*}

The LGRB-SN and ULGRB-SN cases with the data of $z$, $T_{\rm 90}$, $\theta_{\rm j}$, $E_{\rm \gamma,\rm iso}$, $E_{\rm k,\rm iso}$, and $M_{\rm Ni}$ are collected in Table 1. The values of $T_{\rm 90}$ in the table vary from 2.8 to 162.83 seconds, and the redshifts $z$ of most GRBs are less than 1. By comparing the theoretical values of $L_{\rm j,t}$ and $M_{\rm Ni,t}$ provided by the different progenitor models with the observational data $L_{\rm j}$ and $M_{\rm Ni}$, we can constrain the characteristics of progenitor stars of the GRB-SN cases.

In Figure 1, we constrain the masses and metallicities of the progenitor stars of LGRB-SN cases for different outflow rates ($f=50\%$ and $90\%$) and observable activity durations of the LGRB central engine ($t_{\rm a}=10\rm~s$, $30\rm~s$, and $100 \rm~s$). The observational data are denoted by empty black circles. Here, the jet luminosities calculated by the data in Table 1 are just the upper limits. The different colors of the filled symbols represent the metallicity values $Z=Z_{\odot}, 10^{-1}Z_{\odot}, 10^{-2}Z_{\odot}, 10^{-4}Z_{\odot}$, and 0. Symbols with the same color in the sequence of the mass values are connected by lines. The mass is in the range of $16-40~M_{\odot}$ with an interval of $2~M_{\odot}$. The progenitor stars with 38 $M_\odot$ and $Z_{\odot}$ cannot supply the accretion processes lasting approximately 100 s, so we marked all the dimensionless mass values of the stars with $Z_{\odot}$ in Figures (c) and (f).

We found that all the progenitor stars of $M<30~M_{\odot}$ with $Z=10^{-1}Z_{\odot}$ or $Z=Z_{\odot}$ failed to explain all the data in the cases of $f=90\%$. However, most of the LGRB-SN cases can be satisfied with the massive ($M>34~M_{\odot}$) and solar-metallicity stars or the low-metallicity ($Z\lesssim 10^{-2}Z_{\odot}$) stars. In theoretical calculations, the larger outflow rates $f$ result in the more massive $^{56}\rm Ni$ materials and the lower jet luminosity of LGRBs. For $f=90\%$, most of the data require stars with low metallicity and low mass $\lesssim 26~M_\odot$. Furthermore, as seen by comparing Figures 1 (a)-(c) and (d)-(f), the values of $t_{\rm a}$ have not significantly affected the demands for the progenitor stars.

Ultra-LGRBs (ULGRBs) were once considered a new population of GRBs. We proposed that compared with the progenitor stars of LGRBs, these of ULGRBs are not unique \citep{Liu2018b}.

GRB 111209A is reported to be associated with SN 2011kl \citep{Greiner2015}. The SN is more than three times more luminous than the typical type Ic SNe and has become the most luminous GRB-SN detected so far \citep{Kann2016}. The energy source of SN 2011kl is still a mystery. \cite{Greiner2015} obtained the $^{56}\rm Ni$ mass $M_{\rm Ni}=1.0\pm0.1~M_{\odot}$ and the ejecta mass $M_{\rm ej}=3.0\pm1.0~M_{\odot}$. Considering a near-infrared correction, \cite{Kann2016} derived $M_{\rm Ni}=2.27\pm0.64~M_{\odot}$ and $M_{\rm ej}=6.79_{-2.84}^{+3.67}~M_{\odot}$ using a two-component $\rm ^{56}Ni$ decay model. The ratio of $M_{\rm Ni}/ M_{\rm ej}\approx0.3$ is too large compared to the value of $~0.07$ inferred for the general GRB-SN population if one considered that $^{56}\rm Ni$ originates entirely from the explosion \citep{Cano2013}. It is a `crisis' on the $^{56}\rm Ni$ shortage for the traditional SN theories, so someone proposed that this SN could not be powered entirely (or at all) by radioactive heating, and magnetars may be an alternative mechanism. \cite{Nakauchi2013} proposed that the cocoon fireball photospheric emissions can explain the superluminous-SN-like bumps.

In the BH hyperaccretion inflow-outflow model, we use the isotropic energy and $^{56}\rm Ni$ mass to constrain the progenitor stars of GRB 111209A. Similar to Figure 1, the progenitors with the different masses and metallicities are represented in Figure 2. The star masses are in the range of $16-40~M_{\odot}$ with the interval of $2~M_{\odot}$ and mark the numbers in the unit of $M_{\odot}$ for the mass values larger than $40~M_{\odot}$. In these two figures, the stars with $> 32~M_\odot$ and $Z_{\odot}$ cannot supply the accretion processes lasting a longer duration as GRB 111209A. For the strong outflow rate $f=90\%$, it is easy to find that some low-metallicity ($Z\lesssim 10^{-2}Z_{\odot}$) progenitors with $M\gtrsim20 ~M_{\odot}$ can produce this ULGRB-SN event. For the moderate outflow rate $f=50\%$, only the massive ($M\gtrsim50 ~M_{\odot}$) and low-metallicity progenitors can meet the observation requirement. Whatever the cases in our model, there is sufficient $^{56}\rm Ni$ to power the luminous SN considering the effects of the disk outflows on the nucleosynthesis.

\section{Conclusions}\label{Conclusions and Discussion}

In this paper, we focused on LGRBs associated with SNe and explored the characteristics of their progenitor stars. Comparing with the GRB luminosity and $^{56}\rm Ni$ mass derived from the data of 15 GRB-SN cases, we constrain the features of LGRBs and ULGRBs. By considering SNe purely powered by the radioactive decay in the disk outflows and GRB jets produced by the BZ mechanism, we found that LGRB-SNe originate from low-metallicity ($Z\lesssim 10^{-2}Z_{\odot}$) stars or massive solar-metallicity stars. For ULGRBs, GRB 111209A, stars with the solar metallicity and a tenth of solar metallicity failed to satisfy the demand of the jet luminosity and $^{56}\rm Ni$ mass. Most of the low-metallicity ($Z\lesssim 10^{-2}Z_{\odot}$) and massive stars could produce GRB 111209A. There is no crisis on the $^{56}\rm Ni$ shortage for luminous SN 2011kl. Moreover, if the activity timescale of the GRB central engine is longer than $T_{90}$, the results in this paper are not contradictory with those in Paper I.

It is worth noting that all cases in this paper are single-star progenitors. However, the binary stars might also play important roles in producing LGRBs \citep[e.g.,][]{Zhang2001,Podsiadlowski2010}. The evolutions of an isolated massive star and a star in the close binary system are very different \citep[e.g.,][]{Heger2003,Chevalier2012,Qian2018}. \cite{Sana2012} found that more than seventy percent of massive stars will exchange materials with their companion stars and lead to a binary merger in one-third of the cases. Furthermore, not all Ib/c SNe originate from Wolf-Rayet single-stars, which might exist in the binary systems \citep[e.g.,][]{Smith2011,Eldridge2013}.

\end{document}